\documentclass{kapproc} % Computer Modern font calls

\RequirePackage{graphicx}%
\RequirePackage{epsf}%
\input{psfig.sty}

\upperandlowercase
\let\footnote\savefootnote
\let\footnotetext\savefootnotetext

\kluwerbib % will produce  bibliography
\def\hii{H\,{\sc ii}}

\begin{document}

%------------ article title  ------------------->>

\articletitle{The massive star IMF at high metallicity}

%% optional, to supply a subtitle:
%\articlesubtitle{Spineto@50}

%% Supply a shorter version of the title for the running head:
\chaptitlerunninghead{The massive star IMF at high metallicity}

%------ author/affiliation choices -------------->>

%% Single author or several authors with same affiliation

 \author{Fabio Bresolin}
 \affil{Institute for Astronomy, 2680 Woodlawn Drive, Honolulu, USA}
 \email{bresolin@ifa.hawaii.edu}

% abstract
 \begin{abstract}
 The question of the variation of the upper IMF at high
 metallicity is briefly reviewed. I show recent results
 suggesting a revision in the definition of `high metallicity' in
 extragalactic \hii\/ regions. I present preliminary results
 concerning constraints on the upper mass limit in metal-rich
 spiral galaxies derived from the detection of Wolf-Rayet stars in the spectra of their \hii\/
 regions.
 The current evidence is in support of an IMF
 extending up to at least 60--70~$M_\odot$ at an oxygen abundance 1--1.5 times the solar value.
 \end{abstract}

%------------ body of article ------------------->>
\section{Questions asked}
After roughly three decades since the pioneering work on the
massive stellar content of extragalactic \hii\/ regions by
\cite{searle71} and \cite{shields76}, there is still space for
discussions regarding the possible variation of the Initial Mass
Function (IMF) properties at high metallicity. This question has
been commonly investigated spectroscopically via the analysis of
the nebular excitation produced by unresolved populations of
stars, embedded in giant \hii\/ regions located within spiral
galaxies. This method is prone to uncertainties, due to the
model-dependent conclusions one can draw on the shape of the IMF.
Despite the known presence of massive stars in the metal-rich
Galactic center (see, for example, Figer in this volume) and the
lack of evidence for variations of the IMF between the Milky Way
and the comparatively metal-deficient Magellanic Clouds (Massey et
al.~1995), we still need to investigate the possible dependence of
the massive star IMF on additional factors, such as the star
formation history, the stellar density, and the galactic Hubble
type. Moreover, quantifying the chemical abundances in metal-rich
star forming regions of spiral galaxies remains, perhaps somewhat
surprisingly, an open issue.

In a recent review \cite{schaerer03} covered several aspects of
the massive star IMF, including the topic discussed in the current
contribution. I therefore concentrate on the most recent results
and on some of the current work being done on the subject.

\section{A different IMF needed?}
The upper end of the mass function is loosely defined here as that
tail composed by stars more massive than $20\,M_\odot$, i.e.~O and
B stars with effective temperatures above 25,000~K. Although rare,
these stars have an important feedback effect on galactic
evolution, via the energy and momentum transferred to the
interstellar medium by stellar winds, as well as its chemical
enrichment, during their whole lifetime up to the supernova
deflagration finale. The rarity and the short lifetimes (only a
few Myr) of massive stars imply that we must account for
statistical effects in the random sampling of the upper IMF, and
that the stellar ensemble under consideration needs to be very
young.

The notion that stars more massive than a certain threshold do not
form at high metallicity (approximately solar and above) derives
from early observational trends in samples of extragalactic \hii\/
regions, combined with the accretion theory of Kahn (1974). The
radial gradients in excitation, measured from the intensity of
forbidden metal lines, and in the equivalent width of the nebular
H$\beta$ emission line in spiral galaxies led to the suggestion
that the upper mass limit is lowered (Shields \& Tinsley~1976) and
that the slope of the IMF becomes steeper (Terlevich \&
Melnick~1981) at large metallicity. This idea has received support
even recently  from further optical and infrared spectroscopy of
extragalactic \hii\/ regions (among others: Goldader et al.~1997,
Bresolin et al.~1999, Thornley et al.~2000). This interpretation
relies on the observed softening of the radiation field at high
metallicity, seen, for example, from the decreasing He\,{\sc
i}\,$\lambda$5876/H$\beta$ line ratio in the optical (Bresolin et
al.~1999, 2004) and from small fine-structure line ratios
(e.g.~[Ne\,{\sc iii}]/Ne\,[{\sc ii}]) in the mid-IR (Rigby \&
Rieke~2004; see Leitherer in this volume).

By contrast, the UV spectral properties of regions of active star
formation do not support the idea of a varying IMF with
metallicity. In particular, the strengths and P\,Cygni profiles of
wind resonance lines, such as C\,{\sc iv}\,$\lambda$1550 and
Si\,{\sc iv}\,$\lambda$1400, in supposedly metal-rich starbursts
can be modeled with a `normal' Salpeter-slope IMF extending up to
100~$M_\odot$ (Gonz\'alez Delgado et al.~2002). An additional
direct probe for the presence of massive stars, the Wolf-Rayet
(W-R) emission feature at 4660\,\AA, has been used to infer the
extension of the IMF to large masses ($>$\,30--40~$M_\odot$) even
at the highest metallicities sampled (Schaerer et al.~2000,
Bresolin \& Kennicutt~2002, Pindao et al.~2002).

The dichotomy in the IMF properties derived from {\em indirect}
(analysis of nebular lines) and {\em direct} (UV lines, W-R
features) investigation methods  seems now to be, at least in
part, the result of the inadequacy of the stellar atmosphere
models used in the past for the calculation of the ionizing flux
output by hot and massive stars. As shown by Gonz\'alez Delgado et
al.~(2002) and Rigby \& Rieke~(2004), the adoption of recent
non-LTE stellar atmospheres, which include the effects of
line-driven winds and line blocking from metals, into the
evolutionary population synthesis models used for the
interpretation of the spectra, leads to more standard conclusions
regarding the upper IMF at high metallicity. In addition,
complications arising from the effects of the nebular geometry and
density structure of \hii\/ regions conjure to make the
determination of IMF parameters from nebular lines alone uncertain
at best.

\section{What {\em is} metal-rich?}
The determination of chemical compositions is a topic where
nebular lines {\em do} provide an essential insight into the
physical and evolutionary status of star-forming galaxies. Most of
our knowledge of radial abundance gradients in spiral galaxies
derives, in fact, from the analysis of forbidden lines in \hii\/
regions.

\begin{figure}[ht]
\centerline{\psfig{file=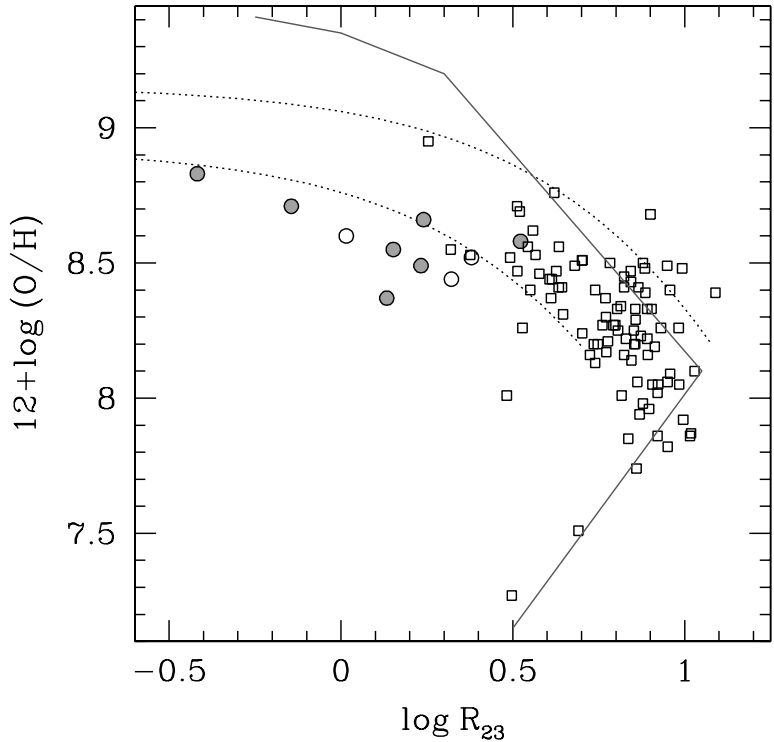,height=4in}}
\caption{Comparison between extragalactic \hii\/ region O/H
abundances measured from electron temperatures (dots) and from the
semi-empirical abundance indicator R$_{23}$. The M51 data by
Bresolin et al.~(2004) are shown as solid circles. Two different
calibrations are used for R$_{23}$: Edmunds \& Pagel 1984 (full
line) and Pilyugin 2001 (dotted lines), the latter for two
representative values of the excitation parameter.}
\end{figure}

The presence of chemical abundance gradients in spiral galaxies is
well-established, but recent extragalactic  nebular abundance work
is questioning the high end of the metallicity scale of previous
investigations. Only recently the faint auroral lines used to
determine direct electron temperatures of the nebular gas have
become observable at high metallicity, where such lines become
extremely faint, requiring large-aperture telescopes for their
detection. In the case of M101, arguably the spiral galaxy with
the best determination of an abundance gradient, Kennicutt et
al.~(2003) found a reduction of the central abundance by up to a
factor of two with respect to indirect methods relying on strong
emission lines (the R$_{23}$ indicator of Pagel et al.~1979). In
the metal-rich spiral M51, the measurement of the auroral lines
[N\,{\sc II}]\,$\lambda$5755 and [S\,{\sc III}]\,$\lambda$6312
from Keck LRIS spectra by Bresolin et al.~(2004) in a significant
number of \hii\/ regions led to the determination of an
extrapolated central abundance  $\log\,$(O/H) = $-3.28$, a roughly
solar value, and a factor up to 2-3 times lower than indicated by
previous investigations. Figure~1 shows how different calibrations
of the R$_{23}$ indicator exceed the abundance inferred from the
electron temperatures, believed to represent the correct value.

This and similar results indicate that the term `high metallicity'
needs to be somewhat revisited when referring to extragalactic
\hii\/ regions. Nebulae that in the past were considered to be of
highly supersolar abundance, are very likely to be in the solar
abundance regime, perhaps up to 50\% higher in the most extreme
cases. The results mentioned earlier concerning the massive IMF in
our own Galaxy and the lack of evidence for observable differences
in the metallicity range bracketed by the Small Magellanic Cloud
and the Milky Way might then imply that no variation in the upper
mass limit is to be expected in the majority of putative
metal-rich star forming galaxies. We might still have to find an
\hii\/ region with 2--3 times the solar oxygen abundance.

\section{Usefulness of W-R features}
The detection and measurement of W-R features (e.g.~the
4660\,\AA\/ `bump') in the spectra of extragalactic \hii\/ regions
at high metallicity, such as in the nucleus of the spiral galaxy
M83 (Bresolin \& Kennicutt~2002), offers a powerful method to
constrain the upper mass cutoff of star-forming regions. This was
elegantly shown by Pindao et al.~(2002), who estimated the {\em
minimum} mass of the most massive stars from model predictions of
the equivalent width of the H$\beta$ emission line, an
evolutionary chronometer for the ionizing stellar clusters, at the
beginning of the W-R phase.

\begin{figure}[ht]
\centerline{\psfig{file=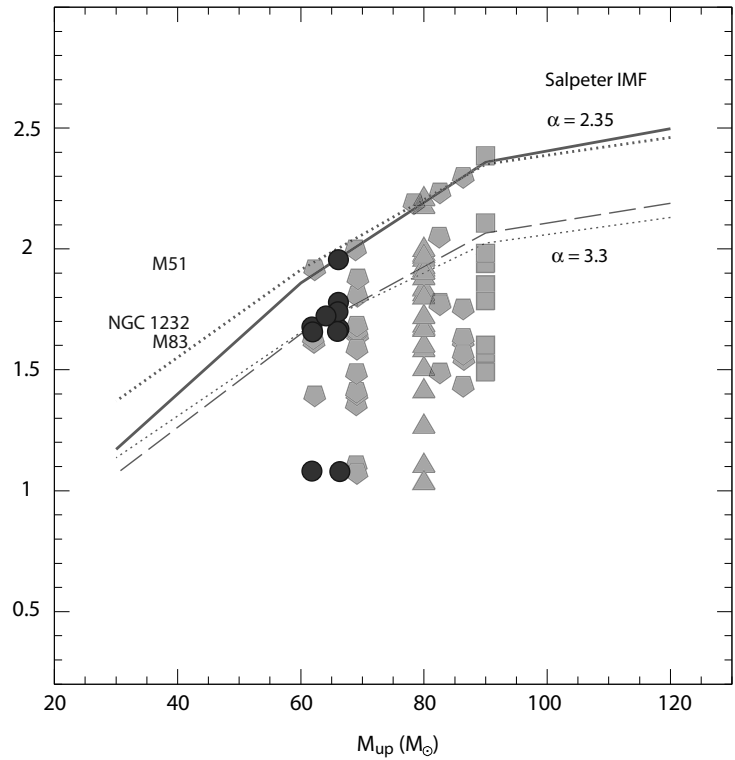,height=4in}}
\caption{Observational data for \hii\/ regions at high metallicity
with W-R features in their spectra are compared to the model
predictions by Schaerer \& Vacca (1998) for two different IMF
slopes, and continuous and burst modes of star formation. By
equating the observed H$\beta$ emission line equivalent width to
its predicted maximum  at the onset of the W-R phase, one obtains
a lower limit for the upper mass limit of the IMF, M$_{\rm up}$.
The dark-grey symbols represent those \hii\/ regions for which we
have reliably determined a solar oxygen abundance or above (in
M51, M83 and NGC~1232) from measurements of their auroral lines.
The light-grey symbols represent supposedly metal-rich objects
from the compilation of Pindao et al.~(2002, squares and
triangles), and remaining \hii\/ regions from our VLT data.}
\end{figure}

We have recently obtained VLT spectra  with the purpose of
analyzing the nebular properties, as well as the massive stellar
(W-R) content, of a sample of extragalactic \hii\/ regions
contained in metal-rich galaxies (Schaerer, Bresolin, Gonz\'alez
Delgado \& Stasi\'nska, in preparation). Some preliminary results
are displayed in Fig.~2, where I consider some of our new
observational data about \hii\/ regions containing W-R features in
NGC~1232, M83 and M51 (dark-grey circles), where the metallicity
is confirmed from electron temperature measurements to be about
solar, or slightly above that, together with the data compiled by
Pindao et al.~(2002, squares and triangles), and the other VLT
targets for which our analysis is still incomplete (remaining
light-grey symbols). Under the minimal assumption that in these
objects the W-R phase just started (which corresponds to a maximum
H$\beta$ equivalent width), the evolutionary models of Schaerer \&
Vacca (1998) tell us that the IMF in these \hii\/ regions extends
up to at least 60--70~$M_\odot$, for a Salpeter-slope IMF. If the
abundance analysis of the remaining objects will confirm their
high metallicity, which is currently just inferred from
strong-line methods, we could push the minimum mass of the upper
mass limit to even higher values, in agreement with the findings
at lower metallicities.

\end{document}